\documentclass[fleqn,twoside]{article}
\usepackage{gc,amsfonts}

\def\d{\mbox{\rm d}}
\def\e{\mbox{\rm e}}

\def\ha{\mbox{$\frac{1}{2}$}}

\def\tqr{\mbox{$\frac{3}{4}$}}

\def\tth{\mbox{$\frac{2}{3}$}}

\def\dddot#1{\mathinner{\buildrel\vbox{\kern5pt\hbox{...}}\over{#1}}}

\def\({\left (}
\def\){\right )}

\def\R{{\mathbb R}}

\heads{S. Cotsakis, G.P. Flessas, P.G.L. Leach and L. Querella}
      {Ever-expanding cosmologies}

\begin{document}
\thispagestyle{empty}
\twocolumn[
%\Arthead{6}{2000}{4 (24)}{1}{12}

\Title
{EVER-EXPANDING, ISOTROPIZING, QUADRATIC COSMOLOGIES\foom 1}

\Authors{S. Cotsakis\foom 2, G.P. Flessas\foom 3, P.G.L. Leach\foom 4}
    {and L. Querella\foom 5}
{GEODYSYC, Department of Mathematics, University of the
    Aegean, Karlovassi 83 200, Greece}
{Groupe de Cosmologie Th\'{e}orique, Institut d'Astrophysique,
    Universit\'{e} de Li\`{e}ge, B--4000 Li\`{e}ge, Belgium}

\Abstract
{We consider some aspects of the global evolution problem of Hamiltonian
homogeneous, anisotropic cosmologies derived from a purely quadratic action
functional of the scalar curvature. We show that models can isotropize in
the positive asymptotic direction and  that quadratic diagonal Bianchi IX
models do not recollapse and may  be regular initially. Although the global
existence and isotropization results we prove hold quite generally, they are
applied to specific Bianchi models in an attempt to describe how certain
dynamical properties uncommon to the general relativity case, become generic
features of these quadratic universes. The question of integrability of the
models is also considered. Our results point to the fact that the more
general models are not integrable in the sense of Painlev\'e and for the
Bianchi IX case this may be connected to the validity of a BKL oscillatory
picture on approach to the singularity in sharp contrast with other higher
order gravity theories that contain an Einstein term and show a monotonic
evolution towards the initial singularity.}

%\vspace{5cm}
% \RAbstract
%     {}
%     {Author(s) in Russian}
%     {Text of abstract in Russian}

] %%%%%%%%%%%%%%

\foox 1 {This work is dedicated to the memory of our beloved
colleague and friend, Professor Jacques Demaret, who passed away
on June 3, 1999.}
\email 2 {skot@aegean.gr}
\email 3 {gfle@aegean.gr}%
\email 4 {leach@math.aegean.gr;\\
    permanent address: School of Mathematical and Statistical Sciences,
    University of Natal, Durban, South Africa 4041}
\email 5 {laurent.querella@hotmail.com}

\section{Introduction}

The problem of describing the past and future asymptotic states of general
cosmological models occupies a central position in mathematical cosmology.
This question can be addressed either in the framework of general relativity
and string theory or in that of higher-order and scalar-tensor theories.
Among several different approaches that have been developed over the years
to tackle the asymptotic problem, two are distinguished, namely, the
Hamiltonian approach and the dynamical systems approach. Both have been and
are being used widely and with great success when dealing with the
asymptotic problem in the framework of general relativistic and string
cosmology but less widely so in the case of higher-order cosmologies, that
is, in the description of the asymptotic states of cosmological models
derived from an action functional which depends nonlinearly on the curvature
invariants.

In this paper we use the Hamiltonian formulation of homogeneous but
anisotropic cosmologies derived from an  action integral purely quadratic in
the scalar curvature (quadratic cosmologies) developed by Demaret and
Querella in \cite{de-qu} to prove several results pertaining to the global
structure of the solution space of these systems. The main motivation for
studying such systems is not so much due to their confrontation with the
currently available empirical data (see, however, \cite{bi}) but to their
different nature.

It is certainly true that recent high-precision tests such as the
Hulse-Taylor binary pulsar (cf.\,\cite{ha-pe}) have confirmed general
relativity to an outstanding degree. This theory predicts that in the case
of the binary pulsar PSR1913+16 the orbit slowly decays and the orbital
period increases due to the loss of energy by emission of gravity waves
and this has  been observed with high precision. However, general relativity
leads to singularities in the spacetimes of all known cosmological models --
places where further predictions about the structure and evolution of the
universe cannot be made. Quadratic curvature terms present in the
gravitational action, which naturally arise in string-theoretic
considerations as quantum gravitational corrections, may rectify this
situation and lead to cosmological models free from such pathologies. One
then envisions the general-relativistic (linear) action to be a low-energy,
low spacetime curvature limit of some quantum gravity theory that would
contain such higher-order curvature invariants and which has yet to be
worked out. Of course the {\em purely} quadratic theories considered here do
not reproduce the weak field limit of general relativity. However, such
theories may gain importance in the high-energy regimes of the early
universe and serve as prototypes from which  more elaborate or natural,
Lagrangian theories of gravity can be built and studied.

Due to their conformal equivalence \cite{ba-co}, the dynamics of
higher-order vacuum cosmologies can be regarded as closely related to that
of general scalar field cosmologies in general relativity.  However, it
should be emphasized that this procedure, although practically useful,
cannot imply a substitute for a full analysis of the original (higher-order)
system since, for example, at those points on the manifold where the
conformal transformation is singular, one cannot obtain a complete picture
of the original dynamics when inverting the conformally related one.
Therefore one has to rely solely on the original (higher order) field
equations when analysing a higher-order system.

We consider vacuum Bianchi cosmologies of class A in the expanding
direction and prove global existence results towards a future
asymptotic state. We show how these quadratic cosmological systems
isotropize and prove that Bianchi IX models in the expanding
direction have a monotonically increasing volume and hence cannot
recollapse. This result is in sharp contrast with the known
behaviour of these systems in general relativity
(cf.\,\cite{li-wa}, \cite{li-wa2}). The question of the algebraic
integrability of quadratic cosmologies is also analysed by
applying the so-called Painlev\'{e} test to the relevant
cosmological equations, and their Lie point symmetries are
calculated.

The plan of this paper is as follows. In the next section we cast the purely
quadratic homogeneous cosmologies in a Hamiltonian form (cf.\,\cite{de-qu})
and write down the basic Hamiltonian cosmological equations. This
Hamiltonian reduction process has an added advantage to that with the usual
Lagrangian field equations, namely, the original fourth-order (Lagrangian)
field equations are reduced to a first order system in the  Hamiltonian
variables.  In \sect 3, by defining new variables, we further reduce the
Hamiltonian equations to a suitable autonomous dynamical system for the
volume and anisotropy functions. \sect 4 is the heart of the paper. We
prove two theorems about the global existence of solutions to the basic
equations in the expanding direction. These theorems are then used in
\sect 5 to prove an isotropization theorem for a class of quadratic,
homogeneous, anisotropic cosmologies, and all previous results are applied
in \sect 6 to several specific Bianchi metrics in an attempt to highlight
their dynamics. In particular, we show, as a corollary of the basic theorems,
that there exist non-recollapsing, ever-expanding Bianchi IX cosmologies in
marked contrast to what is known about such systems in general relativity.
We round up our analysis in \sect 7 by listing the Lie point symmetries of
quadratic cosmologies and performing for them the  Painlev\'{e} test to
decide on their algebraic integrability. We conclude in \sect 8 wherein we
discuss prospects for future work.

\section{Purely quadratic Hamiltonian cosmologies}

In this section we derive the general expressions for the super-Hamiltonian,
supermomenta and canonical equations which are our starting point for the
analysis of relevant cosmological models in subsequent sections.

Our beginning is the purely quadratic action,
\beq
    S=\frac{\beta_{c}}{8}\int \,R^2 (-g)^{1/2}d^{4}x.  \label{action1}
\eeq Following the ADM Hamiltonian prescription, we assume a
foliation of spacetime into spacelike hypersurfaces and use the
ADM coordinate basis to compute the components of various
curvature tensors. The ``0'' index below refers to the normal
component to the hypersurfaces, while the superscripts $^{(4)}$
and $^{(3)}$ indicate quantities defined over the 4-dimensional
spacetime and the spacelike hypersurfaces, respectively. Most of
the time, when there is no ambiguity, the last subscript will be
omitted.  This splitting of spacetime allows one to rewrite the
action (\ref{action1}) in the form \beq
     S=\frac{\beta_{c}}{8}\int N\,^{(4)}\!R^2 \,{}^{(3)}g^{1/2}d^{4}x,
        \label{action2}
\eeq
with
\begin{eqnarray*}
   ^{(4)}\!R \eql ^{(3)}\!R+K^2 -3\tr K^2 -2\frac{\nabla^2 N}{N}\\
            &&\cm - 2g^{kl}{\cal L}_{\vec{n}}K_{kl}, \\
{\cal L}_{\vec{n}}K_{kl}
    \eql \frac{1}{N}(K_{kl,0}-N^{i}K_{kl\mid i}- N_{\ \mid
                k}^{i}K_{il}- N_{\ \mid l}^{i}K_{ik}), \\
K_{ij} \eql  -\frac{1}{2}{\cal L}_{\vec{n}}g_{ij}=-\frac{1}{2N}
                (g_{ij,0}-N_{i\mid j}-N_{j\mid i}),
\end{eqnarray*}
where $K_{ij}$ and $^{(3)}\!R_{ij}$ are the extrinsic and intrinsic
curvature tensors relative to the hypersurfaces mentioned above, the symbol
$\mid $ denotes covariant differentiation on these 3-surfaces, $N$ and
 $N_{i}$ are the lapse and shift functions and ${\cal L}_{\vec{n}}$ denotes
the Lie derivative along the normal $\vec{n}$ to these hypersurfaces.

The Hamiltonian form of (\ref{action2}) is then given, up to
surface terms, by the integral
\beq
    S=\int d^{4}x\,N[ P_{ij}\,{\cal L}_{\vec{n}}Q^{ij}+p^{ij}\,{\cal L}_{
    \vec{n}}g_{ij} - {\cal H}(g,Q,p,P)]          \label{action3}
\eeq
with
\bear {\cal H}(g,Q,p,P)=-2p^{ij}P_{ij}+\frac{Q^2
}{18\beta
        _{c}g^{1/2}}+Q_{\ \ \mid ij}^{ij} \nn
+Q^{ij}R_{ij}  -\frac{Q}{2}R+Q^{ij}P_{ij}P+\frac{Q}{2}(\tr p^2 -p^2 ),
\ear
\bear
Q^{ij} \eql g^{1/2}\frac{\beta _{c}}{2}\;^{(4)}\!R\,g^{ij},
        \nn
p^{ij} \eql -\frac{1}{2}\biggl({\cal L}_{\vec{n}}Q^{ij}-\frac{\delta \tilde{
        {\cal H}}}{\delta K_{ij}}\biggr) ,   \nn
P_{ij} \eql K_{ij},
\ear
where the trace and the traceless part of a tensor $A_{ij}$ are denoted
respectively by $A$ and $A_{ij}^{T}$, $\tr A^2 =A^{ij}A_{ij}$ and
\bearr
\tilde{{\cal H}}(g,Q,K)=KQ^{ij}K_{ij}+Q^{ij}R_{ij}+Q_{\ \ \mid ij}^{ij}\nnn
   \cm -
\frac{Q^2 }{18\beta _{c}g^{1/2}}+\frac{Q}{2}(\tr K^2 - K^2 -R),\label{ham1}
\ear
wherein $Q^{ij}$ is a tensor introduced in order to reduce the order of the
Lagrangian as in Ostrogradski's method. The canonical variables are $g_{ij}$
and $Q^{ij}$ and their conjugate momenta are $p^{ij}$ and $P_{ij}$,
respectively. We can also write the action (\ref{action3}) in the usual
Hamiltonian form, where all the constraints are manifest, namely,
\beq
    S=\int d^{4}x\,\left[ P_{ij}\dot{Q}^{ij}+p^{ij}\dot{g}_{ij} -
        {\cal H}^{*} (g,Q,p,P)\right] ,
\eeq
where we have introduced the total Hamiltonian constraint ${\cal H}^* $ as
\beq
    {\cal H}^* =N{\cal H}+N^{k}{\cal H}_{k}=N^{\mu }{\cal H}_{\mu },
\eeq
${\cal H}$ and ${\cal H}_{k}$ being the super-Hamiltonian and supermomenta
(or spatial constraints) given by (\ref{ham1}) and by
\beq
{\cal H}_{k}=-Q^{ij}P_{ij\mid k}+2(P_{ik}Q^{ij})_{\mid j}-2g_{ik}p_{\ \ \mid
        j}^{ij},
\eeq
respectively.

The 3-metric of any diagonal Bianchi-type model is
\bearr
     d\sigma ^2 = \e^{2\mu }[ \e^{2(\beta _{+}+\sqrt{3}\beta _{-})}(\tilde{
    \omega}^{1})^2 +\e^{2(\beta _{+}-\sqrt{3}\beta _{-})}(\tilde{\omega}
        ^2 )^2 \nnn \inch
    +\e^{-4\beta _{+}}(\tilde{\omega}^{3})^2 ] ,
\ear
where $\mu ,\;\beta _{+},\;\beta _{-}$ are functions of time only and the
set $\{\tilde{\omega}^{i}\}$ is the basis of 1--forms with structure
coefficients $C_{jk}^{i}$,
\[
    d\tilde{\omega}^{i}=\half C_{jk}^{i}\,
        \tilde{\omega}^{j}\wedge \tilde{\omega}^{k}.
\]
Following \cite{de-qu}, we perform a canonical transformation from the
original set of canonical variables $\{g,\,Q,\,p,\,P\}$ to the set
$
\{\mu,\,\beta _{+},\,\beta _{-};\,\Pi _{\mu },\,\Pi _{+},\,\Pi_{-};\,
Q_{+},\,Q_{-},\,Q_{n},$ $P_{+}$, $P_{-}$, $P_{n}\}$, where
\begin{eqnarray*}
g_{11}\eql \e^{2\mu }\e^{2(\beta _{+}+\sqrt{3}\beta _{-})}, \\
g_{22}\eql \e^{2\mu }\e^{2(\beta _{+}-\sqrt{3}\beta _{-})}, \\
g_{33}\eql \e^{2\mu }\e^{-4\beta _{+}},\yy
\Pi_{\ 1}^{1}\eql \frac{1}{12}(2\Pi _{\mu }+\Pi _{+}+\sqrt{3}\Pi _{-}), \\
\Pi _{\ 2}^2 \eql \frac{1}{12}(2\Pi _{\mu }+\Pi _{+}-\sqrt{3}\Pi _{-}), \\
\Pi _{\ 3}^{3}\eql \frac{1}{6}(\Pi _{\mu }-\Pi _{+}), \yy
p^{ij} \eql \Pi ^{ij}+P_{\ \ k}^{Ti}Q^{Tjk}+\frac{Q_{n}}{\sqrt{3}}P^{Tij}+
    \frac{P_{n}}{\sqrt{3}}Q^{Tij}\\
            && \inch + \frac{P_{n}Q_{n}}{\sqrt{3}}g^{ij}, \\
Q^{ij} \eql Q^{Tij}+\frac{Q_{n}}{\sqrt{3}}g^{ij}, \\
P_{ij} \eql P_{ij}^{T}+\frac{P_{n}}{\sqrt{3}}g_{ij}, \\
Q_{\ \ j}^{Ti} \eql \frac{1}{\sqrt{6}}\,\mbox{diag }(Q_{+}+\sqrt{3}
                Q_{-},Q_{+}-\sqrt{3}Q_{-}, -2Q_{+}), \\
P_{\ \ j}^{Ti} \eql \frac{1}{\sqrt{6}} \diag (P_{+}+\sqrt{3}
                P_{-},P_{+}-\sqrt{3}P_{-},-2P_{+}).
\end{eqnarray*}
In terms of these new variables the original Bianchi action can be written as
\begin{eqnarray*}
S \eql \int d^{4}\Omega \, (\Pi _{\mu }\dot{\mu}+\Pi _{+}{\dot{\beta}_{+}}
            +\Pi _{-}{\dot{\beta}_{-}}+P_{+}\dot{Q}_{+}\\
    && \cm +P_{-}\dot{Q}_{-}+P_{n}\dot{Q}_{n}-{\cal H}^* )
\end{eqnarray*}
 with the 4-volume element
$d^{4}\Omega =dt\wedge \tilde{\omega}^{1}\wedge
\tilde{\omega}^2 \wedge \tilde{\omega}^{3}$. The explicit form of
the total Hamiltonian constraint ${\cal H}^* $ depends on the
particular Bianchi model considered. In the case
$L=R^2 $ we have to impose the constraint $Q_{\pm }\approx 0$.
The super-Hamiltonian then reduces to
\bear
    {\cal H}_{R}\eql \frac{\Pi _{+}^2 +\Pi _{-}^2}
            {4\sqrt{3}Q_{n}}-\frac{1}{\sqrt{
              3}}P_{n}\Pi _{\mu }-\frac{2}{\sqrt{3}}P_{n}^2 Q_{n}\nn
&&\ \ +\frac{\e^{-3\mu }}{
    6\beta _{c}}Q_{n}^2 +\frac{\e^{-2\mu }}{\sqrt{3}}Q_{n}V^*,\label{ham2}
\ear
where $V^* = V^* (\beta _{+},\;\beta _{-})$ is directly related to the usual
Bianchi potentials (cf.\,\cite{wa-el}, ch.\,10). Denoting in everything that
follows by $X_y$ the derivative of the function $X$ with respect to the
variable $y$, the canonical equations obtained from (\ref{ham2}) are
\bear
\dot{\mu} \eql -\frac{N}{\sqrt{3}}P_{n}\nn  \dot{\beta}_{\pm
        } \eql \frac{N}{2\sqrt{3}}\frac{\Pi _{\pm }}{Q_{n}}, \nn
\dot{\Pi}_{\mu } \eql N\left[ \frac{1}{2\beta _{c}}\e^{-3\mu }Q_{n}^2
    +\frac{2}{\sqrt{3}}\e^{-2\mu }Q_{n}V^* \right],              \nn
\dot{\Pi}_{\pm } \eql
    -\frac{N}{\sqrt{3}}\e^{-2\mu }Q_{n} V^*_{\beta_{\pm }}, \nn
\dot{Q}_{n} \eql -\frac{N}{\sqrt{3}}(\Pi _{\mu }+4P_{n}Q_{n}), \nn
\dot{P}_{n} \eql \frac{N}{4\sqrt{3}}\left[ \left( \frac{\Pi_{+}}{Q_{n}}
    \right) ^2 +\left( \frac{\Pi_-}{Q_n}\right)^2 +8P_{n}^2 \right]\nn
    &&\ \ -
    \frac{N}{3\beta _{c}}\e^{-3\mu }Q_{n}-\frac{N}{\sqrt{3}}\e^{-2\mu
                }V^*. \label{basic-ham}
\ear

\section{Reduction to an autonomous form}

By imposing the weak constraint ${\cal H}_{R}\approx 0$ and manipulating the
basic Hamiltonian system (\ref{basic-ham}) we obtain
\beq
P_{n}Q_{n}=k-\Pi _{\mu }
\eeq
where $k$ is a constant of integration. When the scalar curvature is not
constant, we can fix the temporal gauge by choosing $^{(4)}\!R=-t$ to find
\bear
    Q_{n} \eql -\frac{\sqrt{3}}{2}\beta _{c}\e^{3\mu }t, \\
    N \eql \frac{3\beta _{c}}{2k}\e^{3\mu }.
\ear
By some further straightforward algebraic manipulations it is possible to
reduce the set of canonical equations to a non-autonomous dynamical system
of three coupled equations for the physical variables $\mu (t),\;\beta _{\pm
}(t)$:
\bear
    \mu ''t+\mu '+A_{1}t^2 \e^{6\mu }-A_2 t\e^{4\mu }V \eql 0,  \nn
    \beta ''\,t+\beta '+A_{3}t\e^{4\mu } V_{\beta} \eql 0,    \label{1}
\ear
where $A_{i}$ are strictly positive constants given by
\beq
    A_{1}=\frac{9\rho }{16},\quad\ A_2 =\frac{3\rho }{2},\quad\
  A_{3}=\frac{3\rho }{8},\quad\ \rho =\frac{\beta_{c}^2 }{k^2 }  \label{1a}
\eeq
with $\beta_{c}^2 $ and $k^2$ arbitrary constants. (The canonical variables
$\mu $ and $\beta _{\pm }$ become uncoupled when the potential term
vanishes, i.e. for Bianchi-type I model.) We can rewrite the dynamical
system (\ref{1}) in the so--called Taub gauge, $d\lambda =\e^{3\mu (t)}dt$,
where $\lambda $ is the physical time with $0\leq \lambda =\lambda (t)\leq
\infty $ and, assuming invertibility, $t=t(\lambda )$, we can write $\mu
(t)=\mu (\lambda )$, $\beta _{+}(t)=$ $\beta _{+}(\lambda )$ or $\beta
_{-}(t)=\beta _{-}(\lambda )$ and $V=V(\beta _{+}(t),\beta _{-}(t))=V(\beta
_{+}(\lambda ),\beta _{-}(\lambda )).$

The treatment of the system of ODEs (\ref{1}) is greatly facilitated by
transforming it into an autonomous system. For this purpose we  introduce
new variables $(t^* ,w,y,z)$ as follows:
\bearr
    \mu (t)=w(t^* ), \cm     \beta _{+}(t)=y(t^* ), \nnn
    \beta_{-}(t)=z(t^* ), \cm        t=\varphi (t^* ),  \label{2}
\ear
where $\varphi$ is a smooth function of $t^* $. On inserting (\ref{2})
into (\ref{1}) we obtain:
\bear
    \frac{\ddot{w}\varphi}{\dot{\varphi }^2 }-\frac{\dot{w}\varphi
        \ddot{\varphi} }{\dot{\varphi }^{3}}+\frac{\dot{w}}
            {\dot{\varphi }}+A_{1}\e^{6w}\varphi ^2 \cm \nn
                -A_2 \e^{4w}\varphi V(y,z)=0,
                        \label{3}\\
    \frac{\ddot{\beta }\varphi}{\dot{\varphi }^2 }- \frac{\dot{
        \beta }\varphi \ddot{\varphi }}{\dot{\varphi }^{3}}+
    \frac{\dot{\beta }}{\dot{\varphi} }+
            A_{3}\e^{4w}V_{\beta}(y,z)\varphi =0,\label{4}
\ear
where the dot denotes differentiation with respect to $t^{*}$ and
\beq
    \beta =y(t^* )\qquad\mbox{\rm or}\qquad z(t^* ).    \label{5}
\eeq
In (\ref{3}) and (\ref{4}) we eliminate $\dot{w}$ and $\dot{\beta }$,
respectively, by requiring that
\beq
    \frac{\varphi \ddot{\varphi} }{\dot{\varphi }^{3}}=
        \frac{1}{\/\dot{\varphi }}          \label{6}
\eeq
and hence
\beq
    \varphi (t^* )=C\e^{k_{1}t^* }.\label{7}
\eeq
\eq (\ref{7}) is the general solution of (\ref{6}) with $C$ and $k_{1}$
arbitrary constants. On the substitution of (\ref{7}) into (\ref{3}) and
(\ref{4}) we deduce that
\bearr
    \frac{\ddot{w}}{Ck_{1}^2 }+ A_{1}C^2 \e^{6w+3k_{1}t^*}
            - A_2 V(y,z)C\e^{4w+2k_{1}t^*}=0, \nnn
    \frac{\ddot{\beta}} {Ck_{1}^2 }+ A_{3}V_{\beta}(y,z)
                    C\e^{4w+2k_{1}t^* }=0 \label{8}
\ear
From (\ref{8}) it is obvious that by introducing a new dependent variable $
x(t^* )$ through
\beq
    w(t^* )=\ha[x(t^* )-k_{1}t^* ]\label{9}
\eeq
and, since it suffices without loss of generality and for our purposes, by
taking in (\ref{7})-(\ref{9})
\beq
    C=k_{1}=-1,                 \label{10}
\eeq
the basic system (\ref{8}) finally assumes the autonomous form
\bear
    \ddot{x}+\alpha_1 \e^{3x}+\alpha_2 \e^{2x}V(y,z)\eql 0,\label{11}\\
    \ddot{y}+\alpha_3\e^{2x} V_{y}(y,z)\eql  0\label{12}\\
    \ddot{z}+\alpha_3\e^{2x}V_{z}(y,z) \eql 0, \label{13}\\
    \mu (t)=\ha [x(t^* )+t^* ],\quad \lal \beta_+ (t)=y(t^*),  \nn
    \beta_-(t)=z(t^* ), \quad \lal t=-\e^{-t^* }        \label{14}
\ear
with $\alpha_1=-2A_1$, $\alpha_2 =-2A_2$, $\alpha_3=A_3$.
All  functions in this system can also be regarded as functions of $\lambda$.

\medskip\noi
{\bf Notation:} In the following we use the two basic time variables $t^*$
and $\lambda$. A prime will denote differentiation with respect to $\lambda$
and an overdot differentiation with respect to $t^*$.

\section{Global existence}

In this section we show that under very plausible hypotheses on the
potential $V(y,z)$, the solutions of the autonomous dynamical system
(\ref{11})-(\ref{13}) can be defined and smoothly extended in the expanding
direction for the whole positive $\lambda$-halfline. The main result of this
section is given in the following two theorems. Theorem 1 basically says
that $t^*$ is a compactified time parameter meaning that its interval of
definition is a compact interval on the real line and Theorem 2 proves that
the interval of definition of the time variable $\lambda$ is the interval
$[0,\infty )$.

\Theorem{Theorem 1} {If the potential
$V=V(\beta_+(\lambda),\beta_-(\lambda))$ is a smooth function of $\beta
_{+}(\lambda )$ and $\beta_-(\lambda )$ and satisfies for all real
$\beta_+(\lambda)$ and $\beta_-(\lambda)$ the condition \beq V=V(\beta
    _{+}(\lambda ),\beta _{-}(\lambda ))\geq M,\quad M\in {\mathbb
                R},     \label{46} \eeq and in the
case $M<0$ the arbitrary constant $\rho >0$ given in (\ref{1a}) fulfils \beq
    \rho <\frac{36(x_0 ')^2 }{27\e^{3x_0 }\,-256M^3},   \label{47}
\eeq
then there exist solutions $\mu (\lambda )$ and $\beta _{+}(\lambda )$,
$\beta _{-}(\lambda )$ to (\ref{1}), which are monotonically increasing and
decreasing, respectively, defined on the interval
$0\leq\lambda\leq\lambda_{\max}$, where
\beq \nq\,
    \lambda_{\max }=\!\int\limits_{-\e^{-t_0^{*}}}^{-\e^{-T^\star}}\exp
    \left\{\frac{3}{2} x [ -\ln (-w)] \right\}(-w)^{-3/2}\d w, \label{48}
\eeq
$T^{\star }$ is a finite number and $x_0 $, $x_0 '$ and $
t_0 ^{\star }$ have been introduced in (\ref{21}). }

\Theorem{Theorem 2} {If Theorem 1 is valid and, in addition, in the case
$\lim_{\lambda \to \lambda _{\max }}\beta _+(\lambda )=-\infty $
and (or) $\lim_{\lambda \to \lambda _{\max }}\beta _-(\lambda )=-\infty $,
the derivatives $V_{\beta}$ are absolutely bounded, that is, for
$n>1$,$k_1 \in \R ^+$, $k_2 \in \R ^+$, $ \beta _+^{(m)}\in \R ^-$,
$\beta_-^{(m)}\in \R^-$, and $\beta_+(\lambda)\in [-\infty,\beta_+^{(m)}]$,
$\beta_-(\lambda )\in [ -\infty, \beta_-^{(m)}]$, the conditions
\bearr
    \left| V_{\beta _+} \beta _+^n \right| <k_1,  \label{49a}\\
    \left| V_{\beta _-} \beta _-^n \right| <k_2   \label{49b}
\ear
hold, then necessarily
\beq
    \lambda _{\max}=\infty, \cm
    \lim_{\lambda \to \infty }\mu (\lambda)=\infty. \label{50}
\eeq
}
We break the proof of these theorems into the two Lemmata below. We define
\beq
    p_1=\dot{x},\qquad p_2 =\dot{y},\qquad p_{3}=\dot{z}. \label{15}
\eeq
Then (\ref{11})--(\ref{13}) become
\bear
    p_{1}p_{1,x}+\alpha_1 \e^{3x}+\alpha _2 \e^{2x}V(y,z)\eql 0,\nn
    p_2 p_{2,y}+\alpha _3 \e^{2x}V_y \eql 0, \nn
    p_{3}p_{3,z}+\alpha _3 \e^{2x}V_z \eql 0,   \label{16}
\ear
and after integration the system (\ref{16}) yields by virtue of (\ref{15}):
\bear
\dot{x}\eql  \biggl(\frac{4}{3} A_{1}\e^{3x}+4A_2 \int
        \e^{2x}V(y,z)dx\biggr)^{1/2},       \label{17}\\
\dot{y}\eql - \biggl(-2A_{3}\int \e^{2x}V_{y} dy\biggr)^{1/2},\label{18}\\
\dot{z}\eql - \biggl(-2A_3\int \e^{2x}V_{z} dz\biggr)^{1/2}.   \label{19}
\ear
Each choice of positive and negative signs in front of the square roots
in (\ref{17}) and (\ref{18}) is taken in order to generate increasing
$x(t^* )$ and decreasing$ \;y(t^* )$ and $z(t^* )$ as functions of $t^* $,
respectively, and  will be justified at the end of this section.  Since the
system (\ref{11})--(\ref{13}) is autonomous, the initial value of $t^* $,
$t_{i}$*, can be arbitrary, and we consider here $t^*_i=t^*_0$.
Therefore, by incorporating into our system (\ref{11})--(\ref{13}) the
physically plausible assumption that $V(y,z)$, $V_{y}(y,z)$ and $V_{z}(y,z)$
are continuous for all $y$ and $z$ in the vicinity of $t^* =t^*_0$, we can
invoke the existence theorem for the dynamical system (\ref{11})--(\ref{13})
and establish the validity in a neighbourhood $\cal {N}$ of $t^* =t_0^*$
of a unique, ${\cal C}^2$ solution $(x(t^*),y(t^*),z(t^*))$,
\bear
    x\eql x(t^*,x_0 ,y_0 ,z_0,\dot{x}_0 ,\dot{y}_0 ,\dot{z}_0), \nn
    y\eql y(t^*,x_0 ,y_0 ,z_0,\dot{x}_0 ,\dot{y}_0 ,\dot{z}_0), \nn
    z\eql z(t^*,x_0 ,y_0 ,z_0,\dot{x}_0 ,\dot{y}_0 ,\dot{z}_0),\label{20}
\ear
satisfying the initial conditions
\bearr
    x(t_0^*) =x_0 ,\qquad \dot{x}(t_0 ^*)=\dot{x}_0 >0, \nnn
    y(t_0^*) =y_0 ,\qquad \dot{y}(t_0 ^*)=\dot{y}_0 <0, \nnn
    z(t_0 ^* )=z_0,\qquad \dot{z}(t_0 ^*)=\dot{z}_0 <0, \label{21}
\ear
where the numbers $x_0 ,,y_0 ,z_0 ,$ are  arbitrary and $\dot{x}_0
,\dot{y}_0$, $\dot{z}_0$ are signed as shown, $x(t^* )$ and $ y(t^* )$,
$z(t^* )$ being monotonically increasing and decreasing functions of $t^* $,
respectively.

Equivalently, due to the monotonicity of $x(t^* )$, we may choose $x$ as the
independent variable and express the solution (\ref{20}) as
\beq
    y\equiv y(x), \cm   z\equiv z(x), \label{22}
\eeq
with the roles of $t^*$ and $x$ interchanged in the first of \eqs
(\ref{20}). Analogous to (\ref{22}) relations can be written down by
considering $y$ or $z$ as independent variables.

Before we consider the existence proof, we proceed to compactify the time
interval $[0,\lambda_{\max}]$. From (\ref{17})--(\ref{19}) we deduce by
using (\ref{21}) and (\ref{22})
\bearr
    T^* -t_0 ^* =\int_{x_0 }^{x_{1}}\,G ^{-1/2}, \nnn
    \cm     x_0 \leq x(t^* )\leq x_{1}=x(T^*)          \label{23}
\ear
where
\bearr
    G(x)=\dot{x}_0^2 +\frac{3}{4}\rho(\e^{3x}-\e^{3x_0}) \nnn
        \cm + 6\rho\int_{x_0}^{x}\e^{2w}V( y(w),z(w)) dw,   \nnn
    T^* -t_0 ^*=
             -\int_{y_0 }^{y_{1}}H^{-1/2}\nnn
    y_{1}=y(T^*)\leq y(t^* )\leq y_0;              \label{24a}
\ear
here \ $H(y)=\dot{y}_0 ^2 +(3/4)\rho\intl_y^{y_0}\e^{2x(w)}V_{y} dw$ \ and
\bearr
     T^* -t_0 ^* =-\int_{z_0 }^{z_{1}}K^{-1/2}, \nnn
        \cm  z_{1}=z(T^* )\leq z(t^* )\leq z_0 ,   \label{24b}
\ear
where $K(z)=\dot{z}_0 ^2 +(3/4)\rho\intl_{z}^{z_0 }\e^{2x(w)}V_{z} dw$.
In (\ref{23}) the upper limit $x_{1}$ in the integral denotes the maximum $
x=x(t^* )$ --- the value for which the solution (\ref{22}) is defined, {\it
i.e.,} $x_0 \leq x\leq x_{1}$, $x_0 \leq w\leq x_{1}$, which implies that
the solution (\ref{20}) holds for $t_0 ^* \leq t^* \leq T^* $ with $x_{1}=$
$x(T^*)$, $y_1 = y(T^*)=y(x_1)$ and $z_1=z(T^*)=z(x_1)$.

We observe now that by virtue of $\d\lambda =\e^{3\mu(t)}\d t$ and (\ref{14})
\beq
    \lambda (t)=\int_{-\e^{-t_0 ^* }}^{t=-\e^{-t^* }}\e^{3\mu
                    (w)}dw \label{25}
\eeq
and that $\lambda _{\max}$ is given precisely by the form in (\ref{48}).
Therefore, by means of (\ref{14}) and (\ref{25}) we have defined a bijective
mapping of the physical interval $0\leq \lambda \leq \lambda _{\max }$ onto
$t_0 ^* \leq t^* \leq T^* $. For this mapping to be physically
legitimate and useful for the treatment of (\ref{1}) we must first prove
(for the class of potentials $V(y,z)$ of interest entering (\ref{1}) or,
equivalently, (\ref{11})--(\ref{13})) that
\beq
        \lambda _{\max }=\infty .\label{26}
\eeq

In conjunction with (\ref{26}) we observe that the integrand in (\ref{25})
can diverge only at $-\e^{-T^{\star }}$. This occurs either for
$\lim_{t^* \to T^* }$ $x(t^* )=\infty $, irrespective of whether $T^* $
is finite of infinite, or for $T^* =\infty $ ($w=0$). In both cases
(\ref{26}) holds.

We first consider the two possibilities regarding the values of $T^* $,
i.e., finite or infinite, and in this respect state the following lemma for
(\ref{23})--(\ref{24b}), which is in fact valid for a class of physically
interesting potentials $V(y,z)$.

\Theorem{Lemma 1} {If the potential $V=V(\beta _{+}(\lambda),\beta
_{-}(\lambda ))$ satisfies for all real $\beta _{+}(\lambda )$ and $\beta
_{-}(\lambda )$ the condition
\beq
    V=V(\beta _{+}(\lambda ),\beta _{-}(\lambda ))\geq M,\cm M\in \R
        \label{27}
\eeq
and in the case $M<0$ the arbitrary constant $\rho >0$ given in
(\ref{1a}) fulfils
\beq
\rho <\frac{36\left( x_0 '\right) ^2 }{27\e^{3x_0 }\,-256M^{3}},\label{28}
\eeq
$x_0 $ and $x_0'$ having been introduced in (\ref{21}), then $T^* $ and
thereby the maximal interval of existence, $[t_0^*$, $T^* ]$, of solution
(\ref{20}) are finite.}

\noi {\bf Proof.} Since by assumption $V\geq M$, $ \,M\in \R ,$ for all real
$\beta _{+}(\lambda ),\beta _{-}(\lambda )$, then due to (\ref{14}) also
$V(y,z)\geq M$, $M\in \R  $, for all real $y$ and $z$, and we obtain the
estimate:
\bearr
G(x) \geq ( \dot{x}_0)^2 +\frac{3}{4}\rho(\e^{3x}-\e^{3x_0 })+6\rho
                M\int\limits_{x_0 }^{x}\e^{2w}\d w  \nnn
 = ( \dot{x}_0)^2 - \frac{3}{4}\rho \e^{3x_0 }-3\rho
    M\e^{2x_0 } + \frac{3}{4}\rho \e^{2x}(\e^{x}+4M), \nnn       \label{29}
\ear
and we call the expression appearing in the last line $F(x)$. Now to further
simplify the notation, let us call $F_{x_0}(x)$ the `value' of $F(x)$ when
the exponential in the fourth term multiplying the round bracket is
$\e^{2x_0}$.  We distinguish the following cases:

{\bf A)} $M+\e^{x_0 }/{4}\geq 0$.

Then, since $x=x(t^* )$ is an increasing function of $t^{\ast
},$ the inequality $M+\e^{x}/4>0$ is valid and, owing to
(\ref{29}), for $x\geq x_0 $
\beq
    F\geq F_{x_0}=\left( \dot{x}_0 \right) ^2 -\frac{3}{4}\rho
        \e^{3x_0 }\ +\frac{3}{4}\rho \e^{2x_0 }\e^{x}>0.\label{30}
\eeq
\eqs (\ref{29}) and (\ref{30}) yield for (\ref{23}) the estimate
\bearr
    0<T^* -t_0 ^*  =\int_{x_0 }^{x_{1}}G^{-1/2}
                \leq \int_{x_0 }^{x_{1}} F^{-1/2}\nnn
    \cm \leq \int_{x_0}^{x_1} F_{x_0}^{-1/2}\,
                    =J_{x_0}(x_{1}).\label{31}
\ear

{\bf B)} $M+\e^{x_0 }/{4}<0\quad  (M < 0)$.

Consider $f(x) = \tqr\rho \e^{2x}(4M+\e^{x})$.

\noi {\bf B$_1$)} $-{8M}/{3}\leq \e^{x_0 }<-4M$. Then $f(x)$ increases for
$x\geq x_0 $.  Thus
\[
 \min F = F(x_0)=(x'_0)^2 >0.
\]
Therefore
\[
F(x) >0,\quad x\geq x_0
\]
 and due to (\ref{29}) we obtain for (\ref{23})
\beq
0\, <\,T^* -t_0 ^* =\int\limits_{x_0 }^{x_{1}}G<
    \int\limits_{x_0 }^{\ln ( -4M)}Fdx+J_{1}(x_{1}),\label{31a}
\eeq
where
\beq
J_{1}(x_{1})=J_{\ln (-4M)}(x_1 ). \label{32a}
\eeq

\noi{\bf B$_2$)} $\e^{x_0 }<-{8M}/{3}.$ Then $f(x) = \tqr\rho
\e^{2x}(4M+\e^{x} )$ decreases for $x_0 \leq x\leq \ln (-{8M}/{3})$ and
attains its minimum value for $x=\ln (-{8M}/{3})$, {\it i.e.,} $\min f(x) =
{64\rho M^{3}}/{9}<0$.  Consequently,
\beq                                 \nq
    \min F=( \dot{x}_0)^2 -3\rho M\e^{2x_0 }-\tqr\rho
        \e^{3x_0 }+\mbox{$\frac{64}{9}$}\rho M^{3}>0 \label{32b}
\eeq
by choosing the arbitrary constant $\rho >0$ in (\ref{32b}) so that
(\ref{28}) holds. Thus we ensure that $F(x)>0,\quad x\geq x_0 $ and yet
again arrive at (\ref{32a}).  Since the integrals $J_{x_0}(\infty )$ and
$J_{1}(\infty )$ in (\ref{31}) and (\ref{32a}), respectively, exist \cite[p.
92, 2.315]{gr}, the lemma is proved.$\DAL$

\medskip
We note that by construction (\ref{23})--(\ref{24b}) as well as the
estimates (\ref{31}) and (\ref{31a}) hold for {\em all} $x_0 ,\dot{x}_0 >0$,
$y_0 $, $\dot{y}_0 <0$ and $z_0 $, $\dot{z}_0 <0$.  Also the condition
(\ref{28}) is not particularly restrictive since it involves a relation
between the arbitrary constants $ \rho >0,$ $x_0 $ and $\dot{x}_0 >0.$

Now, by virtue of Lemma 1 and the discussion following (\ref{26}), we
observe that the remaining possibility for the validity of (\ref{26}) is
that $\lim_{t^* \to T^*}x(t^*)=\infty.$ Consequently the ensuing
developments focus on the proof of this fact for {\em all} $x_0 $,
$\dot{x}_0 >0$, $y_0 $, $\dot{y}_0 <0$ and $z_0 $, $\dot{z}_0 <0$.

To this end we use an immediate and important consequence of Lemma 1, namely
that \cite{hi-sm} {\em at least one} of $\lim_{t^* \to T^* }$ $\left|
x(t^* )\right|$, $\lim_{t^* \to T^* }\left| y(t^* )\right|$, $\lim_{t^* \to
 T^* }\left| z(t^* )\right| $ is $ \infty $.  If for a given $V(y,z)$ the
analytic structure of either one or of both of the quantities ${\partial
 V(y,z)}/{\partial y}$ and ${ \partial V(y,z)}/{\partial z}$ is such as to
exclude in the context of (\ref{24a}) and (\ref{24b}) for all initial
conditions $x_0 $, $\dot{x}_0 >0$, $y_0 $, $\dot{y}_0 <0$ and $z_0 $,
$\dot{z}_0 <0$ the cases $y=-\infty $\ or $z=-\infty $ or $y=-\infty $ {\em
and} $z=-\infty $, then clearly in (\ref{24a}) and (\ref{24b}) $y_{1}\in \R
$ and $z_{1}\in \R $, whence necessarily $ \lim_{t^* \to T^* }$ $x(t^*
)=\infty$.

If, however, the previous cases cannot be as above {\it a priori}
eliminated, upon taking into account the analytic form of some physically
interesting potentials $V(y,z)$, we have the following

\Theorem{Lemma 2} {Suppose that for the initial conditions $x_0 $,
$\dot{x}_0 >0$, $y_0 $, $\dot{y}_0 <0$ and $z_0 $, $\dot{z}_0 <0$,
so that either $y_{1}=-\infty $ or $ z_{1}=-\infty $ or $y=-\infty
$ and $z=-\infty $ appear for either $w\in [-\infty ,y_{m}]$ or $\
w\in \lbrack -\infty ,z_{m}]$, where $y_{m}\in \R^-$ and $z_{m}\in
\R^-$, the derivatives $V_y$, $V_z$ in (\ref{24a}) and (\ref{24b})
are absolutely bounded, that is, for $\in \R ^+,{ }n>1$ \beq
    | V_y w^{n}| \leq k_1 \label{33a}
\eeq
and
\beq
    | V_z w^{n}|\leq k_2  . \label{33b}
\eeq
Then,  for all initial conditions $x_0 $, $\dot{x}_0 >0$,
$y_0 $, $\dot{y}_0 <0$ and $z_0 $, $\dot{z}_0 <0$ we have
\beq
    \lim_{t^* \to T^* }x(t^* )=\infty. \label{34}
\eeq
}

\noi {\bf Proof.} The proof is carried out by {\it reductio ad absurdum}.
Assume firstly the existence of initial conditions $x_0 $, $\dot{x}_0 >0$,
$y_0 $, $\dot{y}_0 <0$ and $z_0 $, $\dot{z}_0 <0$ so that {\em concurrently}
\bearr
    \lim_{w\to y_1=-\infty } x(w)=\lim_{t^* \to T^*} x(t^* )
                =x_1 \in \R ,\nnn
\quad  \lim_{t^* \to T^* } z(t^* ) = z_1 \in \R \qquad\mbox{\rm and}\nnn
    \quad  \lim_{t^* \to T^* }y(t^* )=y_1 = -\infty.
\earn
 We use  (\ref{18}) and (\ref{21}) to obtain
\bear
    \dot{y}\eql -H^{1/2} \label{35a} \\
    \lim_{t^* \to T^* }\dot{y}\eql -H^{1/2}(-\infty ). \label{35b}
\ear
Obviously now for $\ y_{m}<0$, owing to (\ref{33a}) and since $x_0
\leq x(w)\leq x_{1}$ for $ -\infty \leq w\leq y_0 $,  we obtain for the
integral in (\ref{35b})
\bearr
    \biggl| \int_{-\infty }^{y_0 }\e^{2x(w)}V_y\biggr| \leq
        \int_{-\infty }^{y_0 }\left| \e^{2x(w)}V_y \right| dw \nnn
    =\int-{-\infty }^{y_{m}}\left| \e^{2x(w)} V_y
    \right| dw + \int_{y_{m}}^{y_0 }\left| \e^{2x(w)}V_y \right|dw \nnn
    \leq\int_{y_{m}}^{y_0 }\left| \e^{2x(w)}V_y
    \right| dw+ k_{1}\e^{2x_{1}}\int_{-\infty}^{y_{m}}w^{-n}dw.\label{35c}
\ear
Due to the continuity of $V_y$ as a function of $y=y(t^* )$ and $z=z(t^*)
= z(y)$ (owing to the monotonicity of $y(t^*)$), with $-\infty <y(t^* )\leq
y_0 $ $\ $and $z_{1}\leq z(t^* )\leq y_0 $, the integral
\[
    \int_{y_{m}}^{y_0 }\left| \e^{2x(w)}V_y \right| dw
\]
exists. Consequently, the integral on the left-hand side of (\ref{35c}), by
virtue of the estimate on the right-hand side of (\ref{35c}), also
converges. Therefore (\ref{35b}) yields lim$_{t^* \to T^* }\dot{y}$
$=-a_{1}$, $a_{1}\in \R ^{+},$ $a_{1}$ depending of course on $x_0
,\dot{x}_0 >0$, $y_0 ,\dot{y}_0 <0$, $z_0 ,\dot{z}_0 <0.$ This relation
implies that in some left neighbourhood of $T^* $ we have
\[
    \dot{y}=-a (t^* , x_0 ,\dot{x}_0 >0,y_0 ,
        \dot{y}_0 <0, z_0 ,\dot{z}_0 <0)<0,
\]
with $\left| a(t^*, x_0, \dot{x}_0 >0, y_0 ,\dot{y}_0 <0,z_0 ,\dot{z}_0
<0)\right| <\infty $ and $\lim_{t^* \to T^* }a(t^*)=a_{1}$, whereupon
by integration we deduce that in fact $\lim_{t^* \to T^* }y(t^* )$ is finite.

We thus arrive at a contradiction with our hypothesis for the existence of
appropriate $ x_0 $, $\dot{x}_0 >0$, $y_0 $, $\dot{y}_0 <0$ and $z_0 $,
$\dot{z}_0 <0$ such that $\lim_{w\to y_{1}=-\infty }x(w)=\lim_{t^* \to T^*
}x(t^* )=x_{1}\in \R $, $\lim_{t^* \to T^* }z(t^* ) = z_1 \in \R
$ and $\lim_{t^* \to T^* }y(t^* )=y_1=-\infty $.

Likewise, due to (\ref{33b}), it is verified that there do not exist $x_0 $,
$\dot{x}_0 >0$, $y_0 $, $\dot{y}_0 <0$ and $z_0 $, $\dot{z}_0 <0$ such that
$\lim_{w\to z_{1}=-\infty }x(w)=\lim_{t^* \to T^* }x(t^* )=x_{1}\in \R $,
$\lim_{t^* \to T^* }y(t^* )=y_{1}$ and $ \lim_{t^* \to T^* }z(t^*
)=z_{1}=-\infty $ or, owing to (\ref{33a}) and (\ref{33b}), such that
$\lim_{w\to -\infty }x(w)=\lim_{t^* \to T^* }x(t^* )=x_{1}\in \R $,
$\lim_{t^*\to T^* }y(t^* )=y_{1}=-\infty$ and $\lim_{t^* \to T^*} z(t^*
)=-\infty $.

These results, according to the remarks prior to Lemma 2, immediately imply
(\ref{34}). The proof of Lemma 2 is now complete and with this, the proofs
of Theorems 1 and 2 are also complete.$\DAL$

\medskip
At this stage some remarks are appropriate. The conditions (\ref{33a})
and (\ref{33b}) are sufficient only, as we shall later demonstrate, in the
case of the Bianchi IX potential.  Secondly, the validity of (\ref{34}) does
{\em not either } entail $ \lim_{t^* \to T^* }y(t^* )=-\infty $ or
$\lim_{t^* \to T^* }z(t^* )=-\infty $ or both, {\em or } preclude these
possibilities, which may appear for certain values of $ x_0 $, $\dot{x}_0
>0$, $y_0 $, $\dot{y}_0 <0$ and $z_0 $, $\dot{z}_0 <0$ or for specific
analytic forms of $V(y,z)$ ({\it cf. } also the proof of Theorem 3).
Essentially it has been verified that $\lim_{t^* \to T^* }y(t^* )=-\infty $
or $\lim_{t^* \to T^* }z(t^* )=-\infty $ or both are {\em not} possible {\em
without} the validity of \ $\lim_{t^* \to T^* }x(t^* )=\infty .$

On recalling the discussion following (\ref{26}) we deduce that Lemmas 1 and
2 verify (\ref{26}) and, therefore, at least for the class of potentials
$V=V(\beta _{+}(\lambda ),\beta _{-}(\lambda ))=V(y,z)$ satisfying
conditions (\ref{27}), (\ref{33a}) and (\ref{33b}) we have in fact mapped
$0\leq \lambda \leq \lambda _{\max }=\infty $ onto $t_0 ^* \leq t^* \leq
T^{\ast }$, $T^* $ being finite. In this sense we have compactified the
interval $[0,\infty )$ in which $\lambda$ is defined.

To sum up, by utilising the transformations (\ref{14}) and
$\d\lambda =\e^{3\mu (t)}\d t,$ we have
\bear
    \mu '\eql \half (\dot{x}+1) \e^{t^* -3\mu (\lambda )}>0,\label{36}\\
    \beta_{+} '\eql  \dot{y} \e^{t^* }
        \exp \left\{-\fract{3}{2} [x(t^*) + t^*]\right\}
                <0,             \label{37}\\
    \beta_{-}' \eql \dot{z}\e^{t^* }\exp \left\{ -\fract{3}{2}
            [x(t^*)+t^* ]\right\} <0.   \label{38}
\ear
Consequently, by virtue of (\ref{14}), (\ref{17})-(\ref{19}) and (\ref{21}),
(\ref{36})-(\ref{38}) and since $ y(t^* )$ and $z(t^* )$ exist for $t_0 ^*
\leq t^{\ast }\leq T^* $, we conclude that the functions $\mu (\lambda )$
and $\beta _{\pm }(\lambda )$ are defined for $0\leq \lambda <\infty$ and
that they are monotonically increasing and decreasing functions of $\lambda
$, respectively, fulfilling the initial conditions
\bear
    \mu _0 \eql \half (x_0 +t_0 ^* ), \nn
    \mu'_0 \eql \half (\dot{x}_0 +1)
        \exp[t_0 ^* -\fract{3}{2}(x_0 +t_0^*)] > 0, \label{39}\yy
    \beta_{+,0}\eql y_0, \nn
    \beta'_{+,0} \eql \dot{y}_0
            \exp[t_0^* -\fract{3}{2}(x_0 + t_0^*)]<0, \label{40}\yy
    \beta _{-}(0)\eql z_0, \nn
   \beta'_{-,0} \eql \dot{z}_0 \exp
        [t_0^* - \fract{3}{2}(x_0 +t_0^*)]<0,   \label{41}
\ear
whereby, also by virtue of (\ref{23}), (\ref{24a}), (\ref{24b}) and
(\ref{34}),
\bearr
    \lim_{\lambda \to \infty }\mu (\lambda )=\infty ,
            \qquad\mu_0 \leq \mu (\lambda )<\infty, \nnn
    T^* -t_0 ^* =
        \int_{x_0}^{\infty }G^{-1/2},                \nnn
    \lim_{t^* \to T^* } x(t^*)  = x_1 = \infty;
                            \label{42} \\ \lal
 \lim_{\lambda\to \infty} \beta_+ (\lambda) =y_1,
        \qquad  y_1 < \beta _+(\lambda )\leq y_0,  \nnn
    T^* -t_0 ^* =
        \int_{x_0}^{\infty }H^{-1/2},                \nnn
    \lim_{t^* \to T^* }\,y(T^*) =y_1;
                        \label{43} \\ \lal
    \lim_{\lambda\to \infty} \beta_-(\lambda) = z_1,
        \qquad  z_1 <\beta _-(\lambda )\leq z_0,  \nnn
    T^* -t_0 ^* =
            \int_{x_0}^{\infty }K^{-1/2},                \nnn
    \lim_{t^* \to T^* }\,z(T^*) =z_{1}.     \label{44}
\ear
We stress that (\ref{39}) -- (\ref{44}) remain valid for {\em all} $x_0 $,
$\dot{x}_0 >0$, $y_0 $, $\dot{y}_0 <0$ and $z_0 $, $\dot{z}_0 <0$.

\section{Isotropization theorem for quadratic cosmologies}

In view of the results of the previous section, it is now possible to prove
an isotropization theorem for those cosmologies that have potentials
complying with the assumptions of Theorems 1 and 2. We consider the scalar
shear parameter $\Sigma$ and show that it becomes vanishingly small in the
expanding direction. We have

\Theorem{Theorem 3} {The validity of Theorems 1 and 2 implies that the shear
fulfils
\beq
   \lim_{\lambda \to \infty }\Sigma ^2 (\lambda)
    =\lim_{\lambda \to \infty}\left[(\beta _{+}' /\mu ' )^2
+(\beta _{-} ' / \mu ' )^2 \right] = 0.         \label{52}
\eeq
}

\noi{\bf Proof.} We first observe that due to (\ref{36}),
(\ref{37}) and (\ref{38})
\bear
    \beta' _+ /\mu' \eql 2\dot{y}/(\dot{x}+1),     \label{53}\\
    \beta'_- /\mu' \eql 2\dot{z}/(\dot{x}+1),   \label{54}
\ear
with ({\it cf.} (\ref{17}), (\ref{21}))
\beq
    \dot{x}=G^{1/2}.        \label{55}
\eeq
On recalling that $\lim_{w\to y_{1}}\e^{2x(w)}=\infty $, we consider
(\ref{35a}) and observe that the following possibilities exist:
\begin{description}
\item 1) The integral in (\ref{35a}) for $w\to y_{1}$ diverges, in which
case ({\it cf.} (\ref{33a}) for the case $y_{1}=-\infty $ or
$z_{1}=-\infty $ or $ y_{1}=-\infty $ and $z_{1}=-\infty$) at the most
\beq
    \lim_{_{t^* \to T^* }} \dot{y}\sim
            -\lim_{_{t^* \to T^* }} \e^{x(t^* )}. \label{56}
\eeq
\item 2) The integral in (\ref{35a}) for $w\to y_{1}$ converges and thus
\beq
\lim_{t^* \to T^* } \dot{y}=-a, \qquad a\in \R^+.       \label{57}
\eeq
\end{description}
Let it be pointed out that by using standard criteria for the convergence of
integrals one can easily construct examples for possibilities 1) and 2)
above. Cases (\ref{56}) and (\ref{57}), which are of course compatible with
(\ref{33a}) ({\it i.e.,} (\ref{49a})), are the ones possible for the
behaviour of $\lim_{t^* \to T^* }\dot{y},$ in view of $y(t^* )$ being
monotonically decreasing for $t^* \in $ $[$ $t_0 ^* ,T^* ),$ with (\ref{56})
corresponding to $\lim_{t^* \to T^* }y(t^* )=-\infty $ or $\lim_{t^* \to T^*
}y(t^* )=y_{1}\in \R $ and (\ref{57}) relating to $\lim_{t^{\ast }\to T^*
}y(t^* )=y_{1}\in \R $.

Equivalently from (\ref{55}) we deduce that, owing to $V(y,z)\geq
M$, $M\in \R$, at the least
\beq
    \lim_{t^* \to T^* } \dot{x} \sim \lim_{t^* \to T^* }
            \e^{3x(t^* )/2}. \label{58}
\eeq
Therefore (\ref{56})--(\ref{58}) yield for (\ref{53})
\beq
    \lim_{\lambda \to \infty}\frac{\beta_{+}'}{\mu'}=0.\label{59}
\eeq
Along the lines that led to (\ref{59}) it is verified that (\ref{54})
satisfies
\beq
    \lim_{\lambda \to \infty}\frac{\beta_{-}'}{\mu'}=0.\label{60}
\eeq
\eqs (\ref{59}) and (\ref{60}) establish the correctness of (\ref{52}).$\DAL$

\section{Ever-expanding quadratic Bianchi IX cosmologies}

In this section, we apply Theorems 1, 2  and 3 above to some anisotropic
quadratic cosmologies and show as a corollary of the general theory developed
in the previous sections that there exist ever-expanding Bianchi IX
universes in the purely $R^2$ theory, which may also be taken to be regular
initially in sharp contrast with the situation in general relativity.

As a warm-up exercise, let us consider first the easier cases of models of
types I, II and V.

\smallskip

\noi {\bf Case A:} $V=0$ (Bianchi I, $M=0$).

Theorems 1 and 3 trivially hold but Theorem 2 is devoid of any meaning in
the present simple case since, by setting $t_0 ^* =0$ in all the
relevant formulae, without loss of generality we obtain
\bearr
    \lim_{t^* \to T^* } x(t^* ) = \infty, \nnn
    T^* =-\frac{\ln S}{3\sqrt{C_1}},\cm
        S= \frac{\sqrt{A}-\sqrt{C_1}}{\sqrt{A}+\sqrt{C_1}}, \nnn
    C_1=(\dot{x}_0 )^2 - \tqr\rho\, \e^{3x_0} > 0;
                                \label{61}\yy\lal
  \beta_+(\lambda ) = y_0 -\dot{y}_0 \ln [ -t(\lambda)] ,\cm\dot{y}_0 <0;
                \nnn
   \lim_{\lambda \to \infty} \beta _+(\lambda )=y_0 +\dot{y}_0T^*, \nnn
\beta_-(\lambda )
    =z_0 -\dot{z}_0 \ln [ -t(\lambda)] ,\cm \dot{z}_0 <0,  \nnn
\lim_{\lambda \to \infty }\beta _-(\lambda)
                    =z_0 +\dot{z}_0 T^* ,    \label{62}
\ear
where $A=C_{1}+\tqr\rho \e^{3x_0 }$ and $S\in (0,1)$. The case when $C_{1}$
in (\ref{61}) is not positive leads identically to the results of the case
$C_{1}>0$. It is probably interesting to point out that the case $V(y,z)=C=
\const$, which is of no physical importance but fulfils the conditions
for the validity of Theorem 1, yields after a lengthy albeit exact
calculation by means of elliptic integrals the exact finite value of $T^* $
with $\lim_{t^*\to T^* }x(t^* )=\infty $ and, due to (\ref{12})--(\ref{13}),
$\lim_{t^* \to T^* }y(t^* )=y_1\in \R $ and $ \lim_{t^* \to T^* }z(t^* )
=z_{1}\in \R$.

\smallskip
\noi{\bf Case B:} $\ V=\exp[4(\beta _{+}(\lambda )+
    \sqrt{3}\beta_-(\lambda ))]$ (Bianchi II).

\eq (\ref{46}) holds with $M=0$, while (\ref{49a}) and (\ref{49b}) are
satisfied due to the analytic form of $V\ $ and to the fact that $\beta
_{+}(\lambda )$ and $\beta _{-}(\lambda )$ are monotonically decreasing
functions of $\lambda .$ Consequently Theorems 1 to 3 are valid. Note here
that the possibility\ $\lim_{\lambda \to \infty } \beta _{+}(\lambda
)=-\infty $ or  $\lim_{\lambda \to \infty }\beta _{-}(\lambda )=-\infty $ or
both is {\em not\/} excluded.

\smallskip
\noi{\bf Case C:} $V=\e^{4\beta _{+}(\lambda )}$ (Bianchi V).

It is evident that here Theorems 1 to 3 also hold and the possibilty
$\lim_{\lambda \to \infty } \beta _{+}(\lambda )=-\infty $ may appear.

\smallskip
\noi{\bf Case D:} (Bianchi IX).

We now move on to the Bianchi IX case which from the point of view
of dynamics is the most interesting. The usual Bianchi IX potential is
\bear
 V \eql \frac{1}{2}\e^{4\beta _{+}(\lambda )}\left[
    \cosh \left( 4\sqrt{3}\beta _{-}(\lambda )\right) -1\right] \nnn
+ \frac{1}{4}\e^{-8\beta _{+}(\lambda )}\, -\,\e^{-2\beta
    _+(\lambda )}\cosh \left( 2\sqrt{3}\beta _{-}(\lambda )\right).
\earn
Now, on introducing the variables $y(t^* )$ and
$z(t^* )$, by virtue of (\ref{14}) we obtain for the Bianchi IX potential
\beq
    \frac{4V(y,z)}{3}\,+\,1\,=\;V' (y,z),\label{62a}
\eeq
where $V'(y,z)$ is the positive-definite potential used in
\cite{li-wa}. Therefore $M=-3/4$, and Theorem 1 retains its validity for the
Bianchi IX potential. To decide whether Theorems 2 and 3 are valid, we need
\bearr
    V_y = 2\e^{4y}\left[ \cosh \left( 4\sqrt{3}
                z\right) -1\right] -2\e^{-8y}\nnn \inch
        +2\e^{-2y}\cosh (2\sqrt{3}z), \nnn
    V_z = 2\sqrt{3} \Bigl[ \e^{4y}\sinh (4\sqrt{3}z)
        -\e^{-2y}\sinh (2\sqrt{3}z) \Bigr].      \label{63}\nnn
\ear

In the ensuing developments we consider all the possibilities. Unless
otherwise stated, all limits are taken in the direction $y\to -\infty$,
$z\to -\infty$.

{\bf a)} We assume that there exist initial conditions $x_0 ,\dot{x}_0 >0$,
$y_0 ,\dot{y}_0 <0$, $z_0 ,\dot{z}_0 <0$ such that
\bear
    \lim_{t^* \to T^*} x (t^* ) \eql  x_1\in \R ,\nn
    \lim_{t^* \to T^*} y (t^* ) \eql -\infty,  \cm {\rm and}\nn
    \lim_{t^* \to T^*} z (t^* ) \eql z_1 \in \R .
\earn
Then from (\ref{63}) we obtain
\bear
     \lim_{z\to z_1} V_y \eql \lim (-2\e^{8|y|})= -\infty,\nn
     \lim_{z\to z_1} V_z \eql \lim (- \e^{2|y|})= -\infty.  \label{64a}
\ear
From (\ref{64a}a) it is obvious that (\ref{33a}) is violated for the Bianchi
IX potential. Furthermore, (\ref{64a}a) implies, owing to the standard
criterion for the divergence of improper integrals, {\it viz.}
\[
   \lim_{w\to -\infty }(|w|^{\lambda}\e^{8|w|})=\infty,\qquad \lambda \leq 1,
\]
that in (\ref{35b}) the integral tends to $-\infty $ thereby making the
square root complex.  Hence this case cannot occur.

{\bf b)} Suppose now that for appropriate $x_0 ,\dot{x}_0 >0$, $y_0
,\dot{y}_0 <0$, $z_0 ,\dot{z}_0 <0,$ we have $\lim_{t^* \to T^* }x(t^*
)=x_{1}\in \R $, $\lim_{t^* \to T^* }z(t^{\ast })=-\infty$, $\lim_{t^* \to
T^* }y(t^{\ast })=y_{1}\in \R .$ Then (\ref{63}) yields
\bear
  \lim_{y\to y_1} V_y \eql \lim (\e^{4\sqrt{3}|z|})=\infty, \label{65a} \\
  \lim_{y\to y_1} V_z \eql \lim (-\e^{4\sqrt{3}|z|})=-\infty. \label{65b}
\ear
As in possibility a) above, due to the condition (\ref{65b}), (\ref{33b})
ceases to hold for
the Bianchi IX potential.  Moreover, since from (\ref{19}) and (\ref{21}),
\bear
        \dot{z}\eql -K^{1/2},\label{66a} \\
    \lim_{t^* \to T^*}\dot{z}\eql -K^{1/2}(-\infty ),\label{66b}
\ear
we are again led due to (\ref{65b}), so that the square root in
(\ref{66b}) becomes complex. Thus this case is also excluded.

{\bf c)} Finally, let for some specific initial conditions, $x_0 ,\dot{x}_0
>0$, $y_0 ,\dot{y}_0 <0$, $z_0 ,\dot{z}_0 <0$, the result
\[
    \lim_{t^* \to T^* }(x,y,z)=(x_{1},-\infty ,-\infty ) ,
\]
be valid. From (\ref{63}) we deduce
\bear
    \lim V_y \eql \lim (\e^{-4| y| +4\sqrt{3}|z|}
            + \e^{2| y| + 2\sqrt{3}|z|} - 2\e^{8|y|}),\nn
            \nnn\label{67a} \\
    \lim V_z  \eql \lim \sqrt{3}(-\e^{-4|y| +4\sqrt{3}|z|}
                +\e^{2| y| +2\sqrt{3}|z|}).   \nnn\label{67b}
\ear
In the following we  investigate how (\ref{67a}) and (\ref{67b}) impinge on
the validity of (\ref{33a}) and (\ref{33b}) and, consequently, on the
validity of Theorems 2 and 3.

There are various possibilities  to be considered. Firstly, if for
certain initial conditions $x_0 $, $\dot{x}_0 >0$, $y_0 $,
$\dot{y}_0 <0$ and $z_0 $, $\dot{z}_0 <0$, we have
\[
    \lim (|z|/|y|) < \sqrt{3},
\]
then (\ref{67a}) and (\ref{67b}) yield
\beq
    \lim V_y= \lim (-\exp [8|y|])=-\infty   \label{68a}
\eeq
and
\beq
    \lim V_z=\lim\exp[2|y| + 2\sqrt{3}|z|] = \infty, \label{68b}
\eeq
respectively. \eqs (\ref{68a}) and (\ref{68b}) demonstrate that (\ref{33a})
and (\ref{33b}) are not fulfilled. In addition, by virtue of (\ref{68a}) in
(\ref{35b}), the integral tends to $ -\infty$, which renders the square
root complex.  Therefore this case cannot occur. Secondly, suppose that  for
a set of initial conditions, $x_0 $, $\dot{x}_0 >0$, $y_0 $, $\dot{y}_0 <0$
and $z_0 $, $\dot{z}_0 <0$, the relation
\[
    \lim (|z|/|y|) > \sqrt{3}
\]
holds. From (\ref{67a}) and (\ref{67b}) we obtain
\bear
    \lim  V_y =\lim
        \exp\left[ 8|y|+12|y| \left( \frac{|z|}{\sqrt{3}|y|}
         - 1\right)\right] = \infty \label{69a}\nnn
\ear
and
\beq
    \lim V_z = \lim \left[-\exp\left(2|y|+2\sqrt{3}|z|\right)\right]
        =-\infty, \label{69b}
\eeq
respectively. Proceeding as in i) above, we conclude that this possibility
too is excluded. Lastly, if there exist initial conditions, $x_0 $,
$\dot{x}_0 >0$, $y_0 $, $\dot{y}_0 <0$ and $z_0 $, $\dot{z}_0 <0$,
generating
\[
    \lim (|z|/|y|) = \sqrt{3},
\]
(\ref{67a}) and (\ref{67b}) show that
\bear
    \lim V_y  =0, \cm   \lim V_z =0. \label{70}
\ear
From (\ref{70}), (\ref{12}) and (\ref{13}) we infer by integration in the
vicinity of \ $ T^* $, as in the proof of Lemma 2, that $\lim_{t^* \to T^*
}y(t^* )$ and $\lim_{t^* \to T^* }z(t^{\ast }) $ are finite, thus arriving
at a contradiction with the supposition made at the outset of the present
case c).

Cases a) -- c) above immediately show that for the Bianchi IX potential and
for all initial conditions, $x_0 $, $\dot{x}_0 >0$, $y_0 $, $\dot{y}_0 <0$
and $z_0 $, $\dot{z}_0 <0$, we have $\lim_{t^* \to T^*}x(t^*)=\infty$,
$\lim_{t^* \to T^* }y(t^* ) =y_{1}\in \R$ and $\lim_{t^* \to T^* }z(t^* )
=z_{1}\in \R$. We thus arrive at (\ref{34}) {\em without\/} having to resort
to Lemma 2, and precisely such a possibility was announced prior to Lemma 2.
It is straightforward by following the proof of Theorem 3, since now $V_y$
and $V_z$ are bounded for $y\in [y_{1},y_0 ]$, $z\in [z_{1},z_0]$, to
conclude that it holds as (\ref{52}) is based essentially on $\lim_{w\to
y_1}x(w) = \lim_{w\to z_1}x(w)=\lim_{t^*\to T^*}x(t^* )=\infty$. We thus have

\Theorem{Corollary 1} {For the potentials considered in Theorems 1 to 3 we
obtain for the volume $ V_{L}(\lambda )=\e^{2\mu (\lambda )}$ by using
(\ref{36}) that $V_{L}(\lambda )$ is a monotonically increasing function of
$\lambda $, $0\leq \lambda <\infty $ and, owing to (\ref{39}) and
(\ref{50}), fulfils
\bear
    V_{L}(\lambda =0) \eql \exp [2\mu (\lambda=0)]
                    = \exp (x_0 +t_0^* ), \nn
    \lim_{\lambda \to \infty }V_L(\lambda )\eql
    \lim_{\lambda \to \infty }\e^{2\mu (\lambda)} = \infty.\label{71}
\ear }

\noi{\bf Remark.} Had we chosen in (\ref{17}) the minus sign, we would not
have been able to prove the basic Lemma 1. A respective calculation has been
carried out and it does not lead to any sensible results. Further, the plus
sign in either (\ref{18}) or (\ref{19}) or in both would generate increasing
functions $y(t^* )$ or (and) $z(t^* )$, thus preventing us from formulating
(\ref{33a}) or (\ref{33b}) or both in view of the analytic form of the
Bianchi potentials. Finally, it is plausible to use $x(t^*)$ as the
independent variable in (\ref{22}) since it is actually a time variable
according to (\ref{14}).

\section{Lie point symmetries and Painlev\'e analysis of some Bianchi models}

To round off our discussion it is intructive to consider the question of
algebraic integrability of the models considered previously. This is
certainly far from being of purely academic interest since one needs to know
if the known solutions occupy a large part of the phase space or whether
there exist regions where non-integrability manifests itself in a
non-trivial way. In such a case usually one has the phenomenon of the
formation of singular envelopes (cf.\,\cite{co-le}) whereat  any
non-integrable regions are enveloped by the known solutions.

One method to determine the integrability of a system is by a performance of
the so-called {\em singularity, or Painlev\'e, analysis\/} in an effort to
examine whether or not there exists a Laurent expansion of the solution
about a movable pole which contains the number of arbitrary constants
necessary for a general solution. Any other singularities are not permitted
except in the case of branch point singularities which give rise to what is
called the weak Painlev\'e property. A system which is integrable in the
sense of Painlev\'e has its general solution analytic except at the
pole-like singularity (for more details and a nice introduction to the
singularity analysis of dynamical systems we refer the reader to \cite{ta},
ch. 8).

It is interesting that even for these quadratic models where, as we saw
earlier, the global dynamics can be very different from the corresponding
 one in general relativity, integrability is not a common feature in these
examples except for trivial cases.

We first examine the system (\ref{11})--(\ref{13}) for Lie point symmetries.
For the computation we use Program {\tt LIE} due to Head \cite{head1,head2}.

{\bf Case A:}  $V(y,z) = 0$ (Bianchi I), $M=0$.

The equations are particularly simple, being
\bear
\ddot{x} + \alpha_1\e^{3x} = 0\nn
\ddot{y} = 0\nn
\ddot{z} = 0. \label{a1}
\ear
There are nine Lie point symmetries: $\partial_t$, $\partial_y$ ,
$\partial_z$ , $t\partial_y$ , $t\partial_z$,  $y\partial_y$ ,
$z\partial_y$ , $z\partial_y$ , $z\partial_z$. We observe that
most of the symmetry is due to the second and third members of
(\ref{a1}), both of which are trivially integrable. The first has
the solution
\[      \nhq
    x(t) = -\frac{2}{3}\log\left\{\left( \alpha_1 /I\right)^{1/2}\sinh
    \left[\frac{3}{2}\left(2I\right)^{1/2}(t-t_0)\right]\right\},
\]
where $I$ and $t_0$ are constants of integration.

{\bf Case B:} $V = \exp[4(y+\sqrt{3}z)]$, (Bianchi II)

The system is now
\bear
    \ddot{x} - 9K\e^{3x} - 24K\e^{2x+4(y+\sqrt{3}z)} = 0,\nn
    \ddot{y} + 12K\e^{2x+4(y+\sqrt{3}z)} = 0,\nn
    \ddot{z} + 12\sqrt{3}K\e^{2x+4(y+\sqrt{3}z)} = 0.\label{a4}
\ear
Upon introduction of the new variables
\beq
    v = y + \sqrt{3}z,\qquad w = -\sqrt{3}y+z \label{a5}
\eeq
the system (\ref{a4}) takes the simpler form
\bear
    \ddot{x} - 9K\e^{3x} - 24K\e^{2x+4v} \eql 0,\nn
    \ddot{v} + 48K\e^{2x+4v} \eql 0,\nn
    \ddot{w} \eql 0. \label{a6}
\ear
The parameter $K=8\rho$, where $\rho$ was introduced in (\ref{1a}). The Lie
point symmetries of (\ref{a6}) are $\partial_t$ , $\partial_w$ ,
$t\partial_w$ , $w\partial_w$.

We observe a considerable reduction in the number of Lie point symmetries
compared with Case A.  The solution of the third of (\ref{a6}) is trivial,
but of the first two not at all obvious.  There is a first integral
\beq
J = \ha\left(\dot{x}^2-\dot{v}^2\right)-3K\e^{3x}-12K\e^{2x+4v}.\label{a8}
\eeq
\eqs (\ref{a6}a) and (\ref{a6}b) are not in a suitable form for
applying the Painlev\'e test.  Under the transformation
\bear
    r(T) \eql  27K\e^{4(2x+v)},\nn
    s(T) \eql  27K\e^{3x},\nn
    T \eql  \tth i t\label{a9}
\ear
(the constants are chosen for later numerical convenience and do not affect
the essence of the analysis) we obtain the system
\bear
    r\ddot{r}-\dot{r}^2+6r^2s \eql 0,\nn
    s\ddot{s}-\dot{s}^2-s^3+6r \eql 0. \label{a10}
\ear
(We continue to use an overdot to indicate differentiation with respect to
the new time variable, $T$.)  Making the usual substitution
\beq
    r = \alpha\tau^p,\qquad s = \beta\tau^q \label{a11}
\eeq
in (\ref{a10}), we obtain $p=-6, q=-2$ and $p>-6, q=-2$
as the possible exponents.  For the former the leading order behaviour is
\beq
    r = -\ha\tau^{-6},\qquad s = -\tau^{-2}.  \label{a12}
\eeq
When we examine the system (\ref{a10}) for resonances associated with
(\ref{a12}) we find that they are -1, 2 and $\ha(1\pm i\sqrt{35})$.  The
system (\ref{a10}) will not be integrable in the sense of Painlev\'e and,
although the transformation (\ref{a9}) is not homeographic, we can infer the
same for the original pair of equations.  The meagre number of Lie point
symmetries associated with the pair of equations, just $G_1$, reinforces
this conclusion.

{\bf Case C:} $V=\e^{4y}$ (Bianchi V).

The system is now
\bear
    \ddot{x} - 9K\e^{3x} - 24K\e^{2x+4y} \eql 0,\nn
            \ddot{y} + 12K\e^{2x+4y} \eql 0,\nn
                           \ddot{z}  \eql 0. \label{a13}
\ear
The symmetries are just those in Case B above with $w$ replaced with $z$.
The integration of the third equation is trivial and the first two have the
first integral
\beq
    J = \ha\dot{x}^2-2\dot{y}^2-3K\e^{3x}-12K\e^{2x+4y}. \label{a14}
\eeq
As was the case with (\ref{a6}), (\ref{a14}) is not in a form suitable for the
Painlev\'e analysis.  We introduce the transformation
\bear
    r(T) \eql  \e^{3x},\nn
    s(T) \eql  -\e^{2x+4y},\nn
    T \eql i(1/9K)^{1/2}t, \label{a15}
\ear
in which the numerical coefficients are chosen for arithmetrical convenience.
The system (\ref{a14}) becomes
\bear
    r\ddot{r} - \dot{r}^2 + 3r^3 + 5r^2s \eql 0,\nn
            s\ddot{s} - \dot{s}^2 + 2r^2 \eql 0.\label{a16}
\ear
Both indices are $-2$ and the only possible leading order behaviour requires
that all terms be dominant. The resonances are at $-1$, $2$ and $\ha
(1\pm\sqrt{19})$, and we conclude that this system is also not integrable.

{\bf Case D:}  Bianchi IX.

The only symmetry which this system possesses is $\partial_t$, that is, the
system is invariant under time translation.  It is not integrable. It is
known that the non-integrability of this model in general relativity
associated with the BKL oscillatory behaviour of the scale factors in the
contracting direction disappears in the framework of the quadratic theory
derived from the Lagrangian $R+\alpha R^2$  and in all $f(R)$ Lagrangian
gravity theories that contain an Einstein term.  This is due to the
conformal equivalence of such theories to general relativity with a scalar
field matter source and this case is known to have a monotonic approach to
the initial Bianchi IX singularity \cite{ba-co89}. Here, however, we see the
interesting result that in the {\it purely\/} quadratic theory (that is,
without the Einstein term) some sort of non-integrability returns. It
remains to be seen if this lack of integrability of the Bianchi IX spacetime
found here is related to a possible chaotic behaviour in the spacetime
geometry similar to that which occurs in general relativity. If true, such a
result would imply, for instance, that the pure $R^2$ theory is in a sense
closer to general relativity than other quadratic {\it extensions\/} of it
containing the linear Einstein term.

\section{Conclusion}

The behaviour of vacuum quadratic cosmologies in the expanding direction
through the Hamiltonian approach advanced here can be generalized to include
matter fields, and this could be the next problem to tackle in this respect.
How does the inclusion of a scalar field or a perfect fluid affect the
isotropization theorem proved in \sect 5? Is the non-recollapse of the
Bianchi IX model in quadratic gravity stable to matter inclusion or to
perturbations of the purely quadratic scalar curvature Lagrangian?  In
general gelativity all homogeneous, vacuum or matter-filled, closed
cosmologies satisfying the usual energy conditions recollapse
(cf.\,\cite{li-wa}, \cite{li-wa2}). We have shown that the purely quadratic
diagonal Bianchi IX models are ever-expanding (i.e., volume-increasing) and
so they provide a counterexample to a possible closed universe recollapse
conjecture in higher-order gravity theories.

Another question which can be of interest is to analyse these systems in
terms of the expansion-normalized variables of Wainwright (cf. \cite{wa-el})
and compare the results with those arrived at here. This would nicely
complement the isotropization theorem proved in this paper about the scalar
shear variable.

\Acknow{We thank Dr A. Lyberopoulos for drawing our attention to Ref.
\cite{hi-sm} and many useful discussions. PGLL thanks the National Research
Foundation of South Africa and the University of Natal for their continuing
support.}


\small
\begin{thebibliography}{99}

\bibitem{ba-co}
J.D. Barrow and S. Cotsakis, {\it Phys Lett B} {\bf 214}, 515--518
(1988).

\bibitem{ba-co89}
J.D. Barrow and S. Cotsakis, {\it Phys Lett B} {\bf 232}, 172
(1989).

\bibitem{bi}
G.V. Bicknell, {\it J Phys A } {\bf 7},  1061 (1974).

\bibitem{co-le}
S. Cotsakis and P.G.L. Leach, {\it J Phys A} {\bf 27},  1625
(1994).

\bibitem{de-qu}
J. Demaret and L. Querella, {\it Class Quant Grav} {\bf 12},
3085--3101 (1995).

\bibitem{gr}
I.S. Gradshteyn and I.M. Ryzhik, {\it Tables of Integrals, Series
and Products} Alan Jeffrey ed (Academic Press, New York, 1980).

\bibitem{ha-pe}
S.W. Hawking and R. Penrose , {\em The Nature of Space and Time},
(Princeton University Press, Princeton 1996).

\bibitem{head1}
A.K. Head, {\it Comp Phys Commun} {\bf 77}, 241-248 (1993).

\bibitem{hi-sm}
M.W. Hirsch and S. Smale,   {\it Differential Equations, Dynamical
Systems, and Linear Algebra} (Academic Press, San Diego, 1974).

\bibitem{li-wa}
X. Lin  and R.M. Wald, {\it Phys Rev D} {\bf 40},  3280--3286
(1989).

\bibitem{li-wa2}
X. Lin  and R.M. Wald, {\it Phys Rev D} {\bf 4},  2444--2448
(1990).

\bibitem{head2}
J. Sherring, A.K. Head and G.E. Prince,   {\it Math Comp Model}
{\bf 25}, 153-164 (1997).

\bibitem{ta}
M. Tabor,  {\it Chaos and Integrability in Nonlinear Dynamics: An
Introduction} (Academic Press, 1989).

\bibitem{wa-el}
J. Wainwright and G.F.R. Ellis, {\it Dynamical Systems in
Cosmology} (Cambridge University Press, Cambridge, 1997).

\end{thebibliography}
\end{document}